\documentclass[aps,prd,twocolumn,,nofootinbib,showpacs,floatfix]{revtex4}
\usepackage{epsfig}
\usepackage[dvips]{color}

\newcommand{\lesssim}{\; ^< \!\!\!\! _\sim \;}
\newcommand{\gtrsim}{\; ^> \!\!\!\! _\sim \;}
\newcommand{\be}{\begin{equation}}
\newcommand{\ee}{\end{equation}}
\newcommand{\bea}{\begin{eqnarray}}
\newcommand{\eea}{\end{eqnarray}}

\def\alt{\raise0.3ex\hbox{$\;<$\kern-0.75em\raise-1.1ex\hbox{$\sim\;$}}}
\def\agt{\raise0.3ex\hbox{$\;>$\kern-0.75em\raise-1.1ex\hbox{$\sim\;$}}}
\def\phi{\varphi}
\def\sign{{\rm sign}}

\begin{document}

\title{The Galactic magnetic field as spectrograph for ultra-high
  energy cosmic rays}

\author{M.~Kachelrie{\ss}$^{1,2}$, P.~D.~Serpico$^2$, and M. Teshima$^2$}

\affiliation{$^{1}$Institutt for fysikk, NTNU Trondheim, N--7491
  Trondheim, Norway}
\affiliation{$^{2}$Max-Planck-Institut f\"ur
  Physik (Werner-Heisenberg-Institut), D--80805 Munich, Germany}

\begin{abstract}
We study the influence of the regular component of the Galactic
magnetic field (GMF) on the arrival directions of ultra-high energy
cosmic rays (UHECRs). We find that, if the angular resolution of
current experiments has to be fully exploited, deflections in the
GMF cannot be neglected even for $E=10^{20}\,$eV protons, especially
for trajectories along the Galactic plane or crossing the Galactic
center region. On the other hand, the GMF could be used as a
spectrograph to discriminate among different source models and/or
primaries of UHECRs, if its structure would be known with sufficient
precision. We compare several GMF models introduced in the
literature and discuss for the example of the AGASA data set how the
significance of small-scale clustering or correlations with given
astrophysical sources are affected by the GMF. We point out that the
non-uniform exposure to the extragalactic sky induced by the GMF
should be taken into account estimating the significance of
potential (auto-) correlation signals.
\end{abstract}

\pacs{98.70.Sa, 
98.35.Eg   
\hfill MPP-2005-111}

\maketitle

\section{Introduction}

Despite of more than 40~years of research, the field of ultra-high
energy cosmic ray (UHECR) physics poses  still many unsolved
problems~\cite{reviews}. One of the most important open issues is
the question at which energy astronomy with charged particles
becomes possible. The answer depends both on the chemical
composition of UHECRs and on the strength of the Galactic and
extragalactic magnetic fields. Extensive air shower experiments can
in principle measure the chemical composition of the CR flux.
However, the indirect measurement methods and the differences in the
predictions of  hadronic interaction models make the differentiation
between  proton and heavy nuclei primaries a theoretical and
experimental challenge~\cite{chemie}. Other signs for proton or
nuclei primaries are therefore highly desirable.

Complementary information on the charge of the primary may be
obtained by studies of the arrival directions of CRs. Historically,
effects of the geomagnetic field on the CR propagation were decisive
to understand the nature of low-energy CRs: The discovery of the
latitude effect proved that a significant fraction of cosmic rays is
charged, and the east-west asymmetry demonstrated the predominance
of positively charged primaries~\cite{latitude}. It is natural to
ask if the weaker magnetic fields known to exist on larger scales
like the Galactic magnetic field (GMF) might play a similar role at
even higher energies, thus providing important information about the
charge composition and the sources of the UHECRs.

A possible signature of proton primaries is the small-scale
clustering of UHECR arrival directions. The small number of sources
able to accelerate beyond $10^{19}$~eV should result in small-scale
clustering of arrival directions of UHECRs, if deflections in
magnetic fields can be neglected or, to some extent, even in
presence of a magnetized medium (see e.g.~\cite{Takami:2005ij}). For
nuclei with higher electric charge $Ze$, the deflections in the GMF
alone dilute a small-scale clustering signal even at the highest
energies observed. Therefore, the confirmation of the small-scale
clustering observed by the AGASA experiment at energies above
$4\times 10^{19}\:$eV~\cite{Uc00} would favor the hypothesis of
light nuclei primaries, in particular protons. At present, the
statistical significance ascribed to the clustering signal varies
strongly in different analyses~\cite{Uc00,all,Yoshiguchi:2004np}.
Moreover, the HiRes experiment~\cite{Abbasi:2004ib} has not
confirmed clustering yet, but this finding is still compatible with
expectations~\cite{Yoshiguchi:2004np,Kachelriess:2004pc}. The
preliminary data of the Auger Observatory have been searched only
for single sources, with negative result~\cite{Revenu:2005}.\\
Main aim of this work is to quantify the effect of the Galactic
magnetic field (GMF) on the arrival direction of UHECRs and to study
the possibility that corrections for the GMF role may provide
important information about the charge composition and the sources
of the UHECRs.  In particular, the impact of the regular GMF on the
clustering signal and on correlations of UHECRs with BL Lacs is
addressed. Since UHECRs have presumably extragalactic origin, the
possibility of UHECR astronomy relies on the negligible effect of
extragalactic magnetic fields. Existing simulations~\cite{EGMF}
agree on several qualitative features, but disagree on the magnitude
of the UHECR deflections. In the following, we assume optimistically
that such deflections are small in most of the extragalactic sky. In
Sec.~\ref{gmf}, we review the main features of three GMF models
presented previously in the literature. In Sec.~\ref{defl}, we
discuss the role of the GMF for the propagation of UHECRs, and the
method we use to assess the significance of a possible small-scale
clustering in UHECRs data. In Sec.~\ref{AGASAold}, we apply these
concepts to the AGASA data set of events with energy $E\geq 4\times
10^{19}\,$eV, first to autocorrelation and then to test suggested
correlations with BL Lacs. In Sec.~\ref{conc}, we summarize our
results and conclude.

\section{Galactic magnetic field models}\label{gmf}

The first evidence for a Galactic magnetic field was found more than
50 years ago from the observation of linear polarization of
starlight~\cite{hiltner49}. Meanwhile, quite detailed information
about the GMF has been extracted mainly from Faraday rotation
measurements of extragalactic sources or Galactic
pulsars~\cite{Zwiebel1997}. However, at present it is not yet
possible to reconstruct the GMF solely from observations (for an
attempt see~\cite{Stepanov:2001ev}), and instead we will employ
three phenomenological models for the GMF proposed in the
literature. The GMF can be divided into a large-scale regular and a
small-scale turbulent component, with rather different properties
and probably also origin. The root-mean-square deflection
$\delta_{\rm rms}$ of a CR traversing the distance $L$ in a
turbulent field with mean amplitude $B_{\rm rms}$ is
(e.g.~\cite{Harari:2002dy})
\be
 \delta_{\rm rms} = \frac{ZeB_{\rm rms}}{E}\sqrt{\frac{L L_c}{2}}
 \simeq 0.085^{\circ}\frac{Z}{E_{20}} \frac{B_{\rm rms}}{\mu {\rm G}}
        \sqrt{ \frac{L}{\rm kpc} \: \frac{L_c}{50{\rm pc}}}
        \,,\label{deltarms}
\ee
where $L_c$ denotes the coherence length of the field, $E_{20}$ is
the energy in units of $10^{20}$~eV, and $L\gg L_c$ has been
assumed. Recently, Tinyakov and Tkachev noted that the latter
condition could be not fulfilled, at least for some directions in
the sky~\cite{Tinyakov:2004pw}. However, their analysis based
directly on the observed turbulent power spectrum confirmed that the
deflections in the random field are typically one order of magnitude
smaller than those in the regular one. Therefore, we shall neglect the
turbulent component of the GMF in the following.

The regular GMF has a pattern resembling the one of the matter in
the Galaxy and has different properties in the disk and the halo. In
the disk, the field is essentially toroidal, i.e.\ only its radial
($B_r$) and azimuthal ($B_\theta$) components are non-vanishing. The
disk field can be classified according to its symmetry properties
and sign reversals: antisymmetric and symmetric configurations with
respect to the transformation of the azimuthal angle $\theta\to
\theta+\pi$ are called bisymmetrical (BSS) and axisymmetrical (ASS),
respectively. According to the symmetry property with respect to a
reflection at the disk plane ($z\to -z$), the notation A or S is
used: in the first case, the field reverses sign at $z=0$ (odd
field), while in the second case it does not (even field).
Theoretical motivations and observations in external
galaxies~\cite{Sok92} associate the presence of field reversals far
away from the Galactic center (GC) to a BSS geometry: In our Galaxy,
there are probably 3--5 reversals. The closest one is at a distance
of 0.3--0.6 kpc towards the GC, where the higher values seem to be
confirmed from the new wavelet data-analysis used
in~\cite{Stepanov:2001ev,Frick:2000fd}, and about 0.6 kpc is the
value suggested in the review~\cite{Beck:2000dc}. Moreover, there is
increasing evidence for positive $z$ parity (configuration S) of the
GMF near the Sun~\cite{Frick:2000fd,Beck:2000dc,Beck1996}.

\begin{figure}
\begin{center}
\epsfig{file=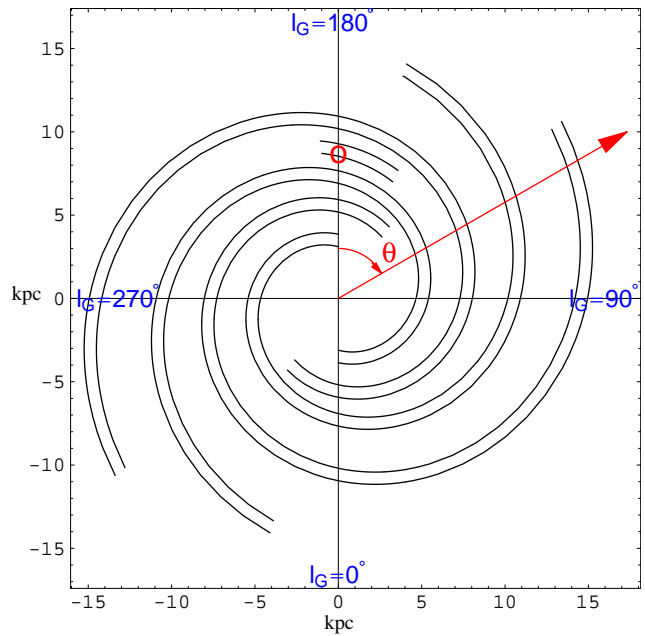,width=1.0\columnwidth} \caption{The
Galactocentric frame used in this paper, together with the Solar
position (circle) along the y-axis and the orientation of the
Galactic polar angle $\theta$. The corresponding Galactic longitudes
are also shown, as well as the Galactic spiral arm model as given
in~\cite{Wainsc92}. \label{figmap}}
\end{center}
\end{figure}

In Galactocentric cylindrical coordinates, the field components in
the disk can be parameterized as
\begin{equation}
 B_r = B(r,\theta)\sin p \,,\qquad
 B_\theta = B(r,\theta)\cos p  \,,
\end{equation}
where $p$ is the pitch angle and $R_0\simeq 8.5$~kpc is the
Galactocentric distance of the Sun, cf. Fig.~\ref{figmap}. Estimates
for the pitch angle are $p=-8^{\circ}\pm 2^{\circ}$ from
pulsar~\cite{Han:2001vf} and starlight polarization data, but other
observations pointing to a value of $p\simeq -13^{\circ}\div
-18^{\circ}$ also exist~\cite{Beck:2000dc}.

The function $B(r,\theta)$ is traditionally modeled reminiscent of
the spiral structure of the matter distribution in the Galaxy as
\begin{equation}
 B(r,\theta)  = b(r)
 \cos\left(\theta-\frac{1}{\tan p}\ln(r/\xi_0)\right) \,.
\label{spiralfield}
\end{equation}
In terms of the distance $d$ to the closest sign reversal, $\xi_0$
can be expressed as $\xi_0 =(R_0+d)\exp(-\frac{\pi}{2}\tan p)$. The
radial profile function $b(r)$ is generally assumed to fall off as
$\propto r^{-1}$~\cite{Stanev97,Tinyakov:2001ir}, consistent with
pulsar measurements \cite{RandLy94}. The behavior of the disk field
in the inner region of the Galaxy is less known, but clearly the
field can not diverge for $r\to 0$. For $r\ge r_{\max}$, the field
is turned off. In the following, we will fix $r_{\max}=20$~kpc. The
vertical profile of the field outside the plane $z=0$ is modeled by
\be
B(r,\theta,z)= f(z) B(r,\theta) \,.
\ee

Despite remaining uncertainties, the regular magnetic field in the
thin disk is yet much better known than other components, namely the
halo (or thick disk) field and a possible dipole field. The first
one could dominate at large Galactic latitudes and the second one
may be of crucial importance near the center of the Galaxy. Because
of the huge volume occupied by the halo field, it may play a
determinant role for UHECR deflections, while the possibly much
higher strength of the field in the center of the Galaxy might
prevent us to access some directions in the UHECR extragalactic sky
(see Sec.~\ref{defl}).

For the halo field, an extrapolation of the thin disk field into the
Galactic halo with a scale height of a few kpc has often been
assumed (e.g.~\cite{Stanev97,Tinyakov:2001ir}). This minimal choice
is in agreement with radio surveys of the thick
disk~\cite{Beuer1985} and mimics the expected behavior of a
``Galactic wind'' diffusing into the halo. However, Faraday rotation
maps~\cite{Han:2001vf,Han1997} of the inner Galaxy
($-90^{\circ}<l_{\rm G}<90^{\circ}$) and of high latitudes ($|b_{\rm
G}|>8^{\circ}$) favor a roughly toroidal component in the halo, of
opposite sign above and below the plane (odd $z$ parity or
configuration A) and with an intensity of
1--2~$\mu$G~\cite{HanMQ1999}. Moreover, there is some evidence for a
$B_z$ component of about $0.2\,\mu$G at the Sun
distance~\cite{Han1994} that could derive from a dipolar structure
at the GC~\cite{Han:2002mz}. In the filaments already detected, the
field strength almost reaches the mG scale~\cite{Beck:2000dc}. Even
if this intriguing picture is roughly consistent with the one
expected if a A0 dynamo mechanism operates in the Galactic halo, it
needs observational confirmation. For example, there is no general
consensus about the existence of such high-intensity magnetic fields
in the central region of the Galaxy, see
e.g.~\cite{Roy:2004sp,LaRosa:2005ai}.

In the following, we review three phenomenological models that
parameterize the regular GMF. These models are characterized by
different symmetries, choices of the functions  $b(r)$ and $f(z)$
and parameters.

\subsection{TT model}\label{TT}

Tinyakov and Tkachev (TT) examined in~\cite{Tinyakov:2001ir} if
correlations of UHECR arrival directions with BL Lacs improve after
correcting for deflections in the GMF. They assumed $b(r)\propto
r^{-1}$ for $r>r_{\min}= 4$~kpc, and $b(r)=$~const. for $r\leq
r_{\min}$. The field $b(r)$ was normalized to 1.4~$\mu$G at the
Solar position. The pitch angle was chosen as $p=-8^\circ$ and the
parameter $d$ fixed to $-0.5$~kpc. They compared a BSS-A and a BSS-S
model and found that for the former model the correlations with BL
Lacs increased. This model has an exponential suppression law,
\be f(z)= \sign(z)\exp(-|z|/z_0) \,,
\ee
with $z_0=1.5$~kpc chosen as a typical halo size. No dipole component
was assumed.

\subsection{HMR model}\label{HMR}

Harari, Mollerach and Roulet (HMR) used in~\cite{Harari:1999it} a
BSS-S model with cosh profiles for both the disk and the halo field
with scale heights of $z_1=0.3$~kpc and $z_2=4$~kpc, respectively,
\be
f(z)=\frac{1}{2\cosh(z/z_1)}+\frac{1}{2\cosh(z/z_2)}\,.
\ee
Thus the disk and halo field share the same spiral-like geometrical
pattern. The function $b(r)$ was chosen as $b(r)= 3R_0/r
\tanh^3(r/r_1)\,\mu$G with $r_1=2$~kpc, hence reducing to
$b(r)\propto r^{-1}$ for $r\gg r_1$ while vanishing at the Galactic
center. The pitch angle was fixed to $p=-10^\circ$, and
$\xi_0=10.55$~kpc. This model represents a slightly modified and
``smoothed'' version of the BSS model discussed by Stanev
in~\cite{Stanev97}. Apart for the vertical profile $f(z)$, the main
differences with respect to the TT model are the $z$ parity and the
$r\to 0$ behavior of the field.

\subsection{PS model}\label{PS}

In~\cite{Prouza:2003yf}, Prouza and Smida (PS) used for the disk
field the same BSS-S configuration as~\cite{Stanev97}, with a single
exponential scale height $z_0$ and $b(r)$ as described in
Sec.~\ref{TT}. In the slightly modified version we use here, we fix
$z_0=0.2\,$kpc, $p=-8^{\circ}$, $d=-0.5\,$kpc and normalize the
local field to $2\,\mu$G. Apart for the larger field-strength, the
main difference with the TT model is the parity of the disk field,
which we take here to be even as in~\cite{Prouza:2003yf}.
Additionally,  we consider a toroidal thick disk/halo contribution,
\bea
B_{Tx}&=&-B_T\:\sign(z)\cos\theta\,,\nonumber \\
B_{Ty}&=&B_T\:\sign(z)\sin\theta\,,
\label{toroidalcomp}
\eea
where
\begin{equation}
B_T=\frac{B_{T,\max}(r)}{1+\left(\frac{|z|-h_T}{w_T}\right)^2} \,,
\end{equation}
$h_T=1.5$~kpc is the height of the maximum above the plane and
$w_T=0.3$~kpc is its lorentzian width. In contrast
to~\cite{Prouza:2003yf}, we choose
\begin{equation}
 B_{T,{\max}}(r) = 1.5\mu {\rm G}
 \left[\Theta(R_0-r)+\Theta(r-R_0)e^{\frac{R_0-r}{R_0}}\right] \,,
\end{equation}
with $\Theta$ denoting the Heaviside step function, so that the halo
contribution becomes  negligible for $r\gg R_0$,
because there is no evidence for such a field outside the solar
circle~\cite{Han1997}. Finally, a dipole field is added as e.g.
in~\cite{Prouza:2003yf,Yoshiguchi:2003mc}, \bea
B_x&=&-3\mu_G\cos\phi\sin\phi\sin\theta/R^3\,,\nonumber\\
B_y&=&-3\mu_G\cos\phi\sin\phi\cos\theta/R^3\,,\nonumber\\
B_z&=&\mu_G(1-3\cos^2\phi)/R^3\,,
\label{dipolefield}
\eea
where $R\equiv\sqrt{r^2+z^2}=\sqrt{x^2+y^2+z^2}$, $\cos\phi\equiv
z/R$ and $\mu_G$ is the magnetic moment of the Galactic dipole with
$\mu_G=123 \mu {\rm G\, kpc}^3$ in order to reproduce $B_z\simeq +
0.2 \mu$G  near the Solar system~\cite{Han1994}. To avoid a
singularity in the center, we set $B_z=-100\,\mu$G inside a sphere
of 500~pc radius centered at the GC. Note that
in~\cite{Prouza:2003yf} values as large as 1~mG were used for the
hard core of the dipolar field. However, data from low frequency
non-thermal radio emissions of electrons~\cite{LaRosa:2005ai} favor
a value of about $10\,\mu$G down to a $10\,$pc scale, and put the
safe bound of $100\,\mu$G which we actually use.

Finally, we warn the reader that these models intend to provide only
a rough approximation to the true structure of the GMF. Moreover,
the GMF models are not self-consistent because the condition
$\nabla\cdot{\bf B}=0$ is not fulfilled by any of the disk fields
discussed: using the ansatz of Eq.~(\ref{spiralfield}) for
$B(r,\theta)$, $\nabla\cdot{\bf B}=0$ can only be satisfied by
$b(r)={\rm const.}$ or a non-vanishing $z$-component of the {\em
disk\/} field.

\section{GMF and UHECR propagation}
\label{defl}
A generalized version of the Liouville theorem was shown to be valid
for CRs propagating in magnetic fields soon after the discovery of
the geomagnetic effect~\cite{33}. For UHECRs, this theorem has been
numerically tested e.g. in~\cite{Alvarez-Muniz:2001vf}, where
particles were injected isotropically from randomly distributed
sources at different Galactocentric distances. Even after the
propagation in the GMF, the sky on Earth appears isotropic (see
their Fig.~6, left panel), while the effective exposure of an
experiment to the extragalactic sky is strongly modified by the GMF
(see their Fig.~6, right panel). Indeed, whenever the rigidity
cutoff can be neglected for high energy cosmic rays, for an
isotropic flux outside the Galaxy the GMF introduces no anisotropy.
At sufficiently low energy, however, the GMF may introduce blind
regions on the external sky, which translate into observed
anisotropies for an Earth-based observer. This effect is easy to
estimate in the simple case in which our Galaxy has a large scale
dipolar field as the one introduced in the PS model. The St{\o}rmer
theory (see e.g.~\cite{latitude}) can be applied to determine the
rigidity cutoff ${\cal R}_{S}$ below which no particle can reach the
Earth. Since the Earth is at zero Galacto-magnetic latitude, we
obtain
\begin{equation} {\cal R
}_{S}=\frac{\mu_G}{2R_0^2}\frac{1}{[1+\sqrt{1-\sin\epsilon}]^2}\,.
\end{equation}
Here, ${\cal R }_{S}$ depends on the arrival direction of the cosmic
ray via the function $\epsilon$, that we do not need to specify
here. Assuming that the tiny vertical component detected at the
solar system of $0.2\,\mu$G is due to a dipole field, we get
$\mu_G\simeq 120\,\mu{\rm G\, kpc}^3$ and then ${\cal R }_{S}$ would
vary in the range $10^{17}\,$V--$10^{18}\,$V. For such a GMF,
large-scale anisotropies should be seen around $E\sim Ze\,{\cal R
}_{S}$, if an extragalactic component dominates at this energy.
Models that invoke a dominating extragalactic proton component
already at $E\simeq 4\times 10^{17}$~eV (see
e.g.~\cite{Ahlers:2005sn}) or extragalactic iron nuclei at $E\alt
10^{19}$~eV might be in conflict with the observed isotropy of the
CR flux. Unfortunately, the previous conclusion is not very robust
against changes in the GMF model. Indeed, the geometry of the GMF is
more complicated than a simple dipole. A naive estimates of the
rigidity cutoff ${\cal R}_{S}$ might be obtained from the size of
the Larmor radius in the Galaxy,
\begin{equation}
r_L = \frac{p}{Ze B}\simeq\frac{{\cal R}}{3\times 10^{15}{\rm V}}\:
      \frac{\mu{\rm G}}{B} \: {\rm pc}.\label{larmorR}
\end{equation}
For $B\simeq$ few $\mu$G and a Galactic magnetic disk of thickness
${\cal O}$(100)~pc, Eq.~(\ref{larmorR}) implies ${\cal R}\lesssim
{\rm few\:}\times 10^{17}$~V.  However, the ${\cal O}$(100)~pc disk
thickness is comparable to the largest scale of turbulent fields.
Thus, quasi-diffusive processes might work in an opposite direction,
partially erasing anisotropies due to blind region effects in an
extragalactic dominated flux. One should also keep in mind that in
the same energy range a sizeable anisotropy is expected in Galactic
models of the origin of CRs, although usually with a different large
scale pattern~\cite{Candia03}. Given present knowledge of the GMF,
we conclude that the signature of blind regions for an extragalactic
flux at ${\cal R}\lesssim {\rm few\:}\times 10^{17}$~V is less
robust and unambiguous than naively estimated.

The previous discussion referred to the case of an incoming
isotropic flux. However, if the AGASA small-scale clusters are not
just a statistical fluctuation, the UHECR flux is, at least on
small-scales, anisotropic. In this case, the CR flux can be
magnified or de-magnified by magnetic lensing phenomena, and the
application of the Liouville theorem is
non-trivial~\cite{Harari:2000az}. The magnification effects of the
GMF change the experimental exposure and a well-defined procedure is
needed to assess the significance of any detected anisotropy. One
can test the significance of observed anisotropies by comparing the
values of the statistical estimator based on the $N_d$ data with a
large number $\cal{N}$ of simulated $N_d$-points samples of the
null-hypothesis. For each set, one should consider the propagation
in the GMF, convolve with the experimental exposure, and finally
reject the null-hypothesis with a given significance. In the
following, we use  for practical reasons the backtracking
technique~\cite{Karakula71}.
\begin{figure}[!htb]
\begin{center}
\epsfig{file=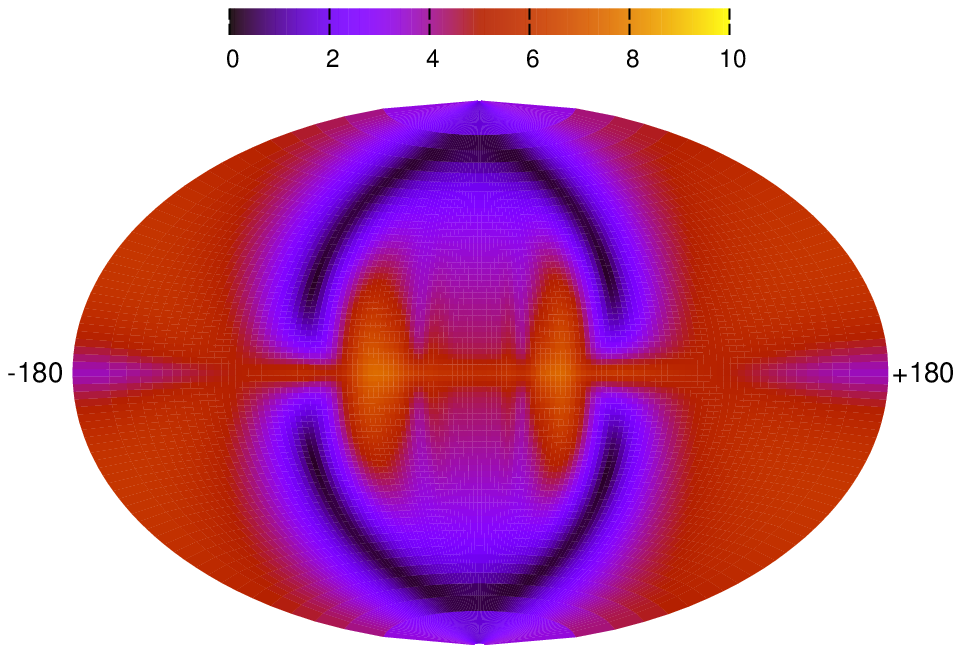,width=0.9\columnwidth}
\epsfig{file=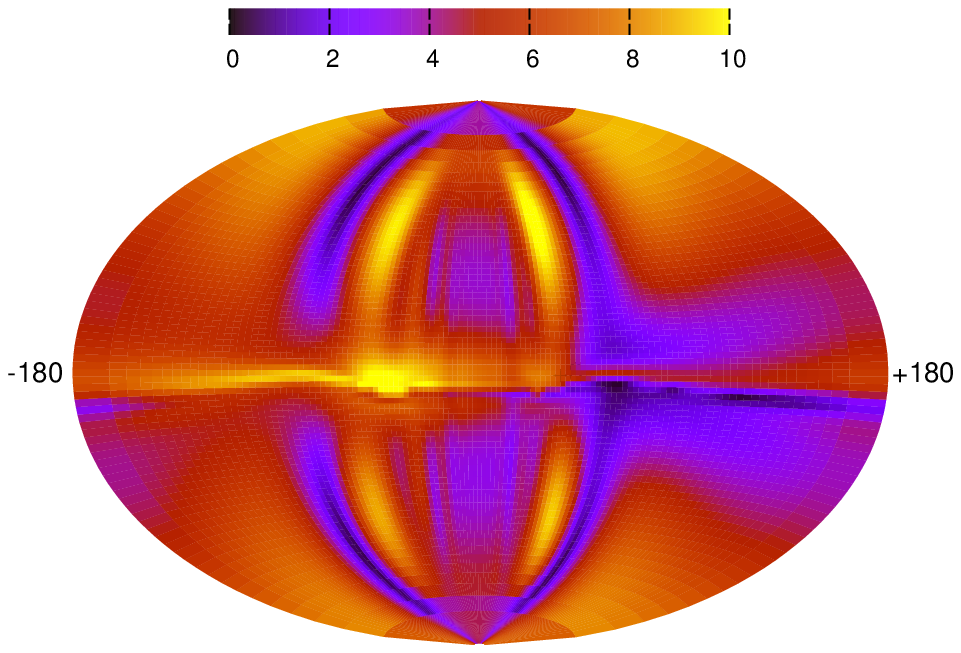,width=0.9\columnwidth}
\epsfig{file=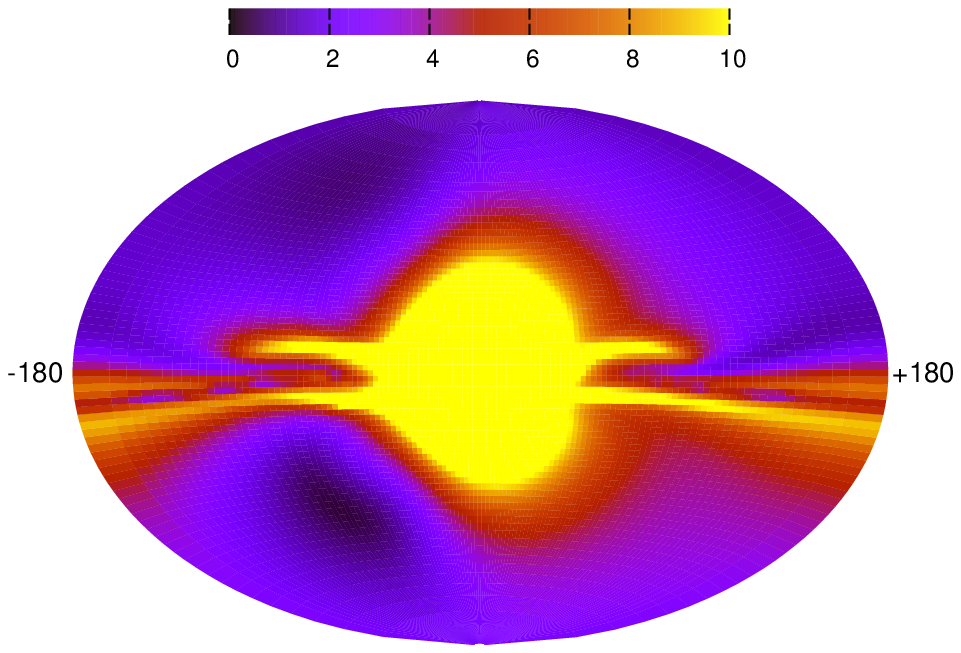,width=0.9\columnwidth}
\caption{Deflection maps for the TT (top), HMR (middle), and PS
(bottom) models of the GMF, for a rigidity of 4$\times$ 10$^{19}$~V.
The deflection scale is in degree, and the maps refer to the
direction {\it as observed at the Earth.\/} All maps use a
Hammer-Aitoff projection of Galactic coordinates. \label{fig:defl}}
\end{center}
\end{figure}
Since we deal with ultra-relativistic particles, the equations of
the motion can be written in the form
\be \frac{d {\bf v}}{dt}=\frac{{\bf v}\times {\bf B}}{\cal{R}} \,,
\label{trajectory} \ee
where ${\bf v}$ is a vector of modulus practically equal to $c=1$.
The integration is stopped when the particle reaches a distance of
50 kpc from the GC. Note that the energy losses of UHECRs on
Galactic scales ($\sim$ 10 kpc) are negligible, provided the
trajectory is not very far from a rectilinear one. In
Fig.~\ref{fig:defl}, we show the map thus obtained, for the three
models considered and a rigidity of $4\times 10^{19}$~V. Note that
the deflection $\delta$ of a particle of rigidity $E/Ze$ in a field
of strength $B$ coherent over the scale $L$ is approximately given
by
\be \delta \simeq 0.53^{\circ}\frac{Z}{E_{20}}
              \frac{B}{\mu {\rm G}}\frac{L}{\rm
              kpc}\,.\label{regdeflEq}
\ee

To obtain an estimate for the average deflection of CRs in different
sky regions, labeled as in Table~\ref{tableregions}, we have
followed backwards 50000 randomly chosen CRs of rigidity ${\cal
R}=10^{20}$~V, for which the hypothesis of a quasi-rectilinear
trajectory is well fulfilled.  In Table~\ref{table1}, we show the
average value and the square root of the variance of ${\cal
R}_{20}\delta$ (i.e., in units of $10^{20}$~V) for the three models
of the regular GMF discussed in the Section~\ref{gmf}, calculated
separately for eight different sky regions. The quantity ${\cal
R}\delta$ depends only on the GMF model and scales almost linearly
with the field strength $B$.
\begin{table*}[!htb]
\begin{center}
\begin{tabular}{|c|c|c|c|c|}
\hline $b_G\backslash l_G$  & 315$\,\le\,l_G\,<\,$45 & 45$\,\le l_G<\,$135 & 135$\,\le l_G<\,$225 &  225$\,\le l_G<\,$315\\
\hline
$|b_G|\ge\,$30  & Ah & Bh & Ch &  Dh\\
\hline
$|b_G|<\,$30  & Al & Bl & Cl &  Dl\\
\hline
\end{tabular}
\caption{The labels used for the eight different regions of the sky
referred to in the text. Angles are in degrees.
  \label{tableregions}}
\end{center}
\end{table*}
The largest difference between the three GMF models occurs in the
region Al: in the PS model, the only one with a dipole field, huge
deflections arise close to the GC. In the regions Bh, Ch, and Dh the
stronger halo fields of the TT and especially the HMR model cause
larger deflections than in the PS model.
In the l-regions, apart for Al, the deflections of the three
models are all of the order $1^{\circ}$--$2^{\circ}$, and
comparable to each other within $1\,\sigma$. Since the CR in these
directions mainly travel through the disk, in order to escape the
Galaxy they  have to cross the regions where the field geometry
and intensity is better known, and a better agreement among the
models exists.

\begin{table*}[!htb]
\begin{center}
\begin{tabular}{|c|c|c|c|}
\hline region  & ${\cal R}_{20}\,\delta$ (PS) & ${\cal R}_{20}\,\delta$ (TT)& ${\cal R}_{20}\,\delta$ (HMR) \\
\hline Ah  & $1.8^\circ\pm 1.0^\circ$ & $0.9^\circ\pm 0.5^\circ$ & $2.1^\circ\pm 0.8^\circ$ \\
\hline Bh  & $1.3^\circ\pm 0.4^\circ$ & $1.3^\circ\pm 0.6^\circ$ & $2.2^\circ\pm 0.8^\circ$ \\
\hline Ch  & $ 0.9^\circ \pm 0.4^\circ$ & $2.0^\circ\pm 0.3^\circ$ & $2.7^\circ\pm 0.5^\circ$\\
\hline Dh  & $0.5^\circ\pm 0.2^\circ$ & $1.1^\circ\pm 0.6^\circ$ & $2.1^\circ\pm 0.9^\circ$ \\
\hline Al  & $14^\circ\pm 21^\circ$ & $1.9^\circ\pm 0.4^\circ$ & $2.2^\circ\pm 0.7^\circ$ \\
\hline Bl  & $2.0^\circ\pm 0.9^\circ$ & $1.7^\circ\pm 0.5^\circ$ & $1.2^\circ\pm 0.4^\circ$ \\
\hline Cl  & $1.7^\circ\pm 1.1^\circ$ & $1.9^\circ\pm 0.5^\circ$ & $1.8^\circ\pm 0.3^\circ$ \\
\hline Dl  & $1.6^\circ\pm 1.0^\circ$ & $1.6^\circ\pm 0.5^\circ$& $2.3^\circ\pm 0.6^\circ$ \\
\hline
\end{tabular}
\caption{The rigidity times average deflections ${\cal
R}_{20}\,\delta$ in the
  eight different regions labeled in Table~\ref{tableregions}. \label{table1}}
\end{center}
\end{table*}

If one excludes the central regions of the Galaxy, the average
deflection is $\delta \simeq 2^{\circ}/{\cal R}_{20}$, and the
differences for the magnitude of the deflections are of the order of
50\% among the models. Thus only for the highest energy events and
proton primaries the role of the GMF is negligible compared to the
angular resolution $\delta_{\rm exp}$ of CR experiments. The latter
is as good as $\delta_{\rm exp}\simeq 0.6^\circ$ for the HiRes
experiment~\cite{Abbasi:2004ib} and for Auger hybrid
events~\cite{AUGERres}. For lower ${\cal R}$, correcting for
deflections in the GMF would becomes crucial to exploit fully the
angular resolution of UHECR experiments. Note also that, even in the
ideal case of perfectly known GMF, a reconstruction of the original
arrival directions would require a relatively good energy
resolution: an uncertainty of, say, 30\% in the energy scale around
$5\times 10^{19}\:$eV would lead to errors $\gtrsim 1^{\circ}$ in
the reconstructed position of proton primaries in most of the sky.

\begin{figure}[!htb]
\begin{center}
\epsfig{file=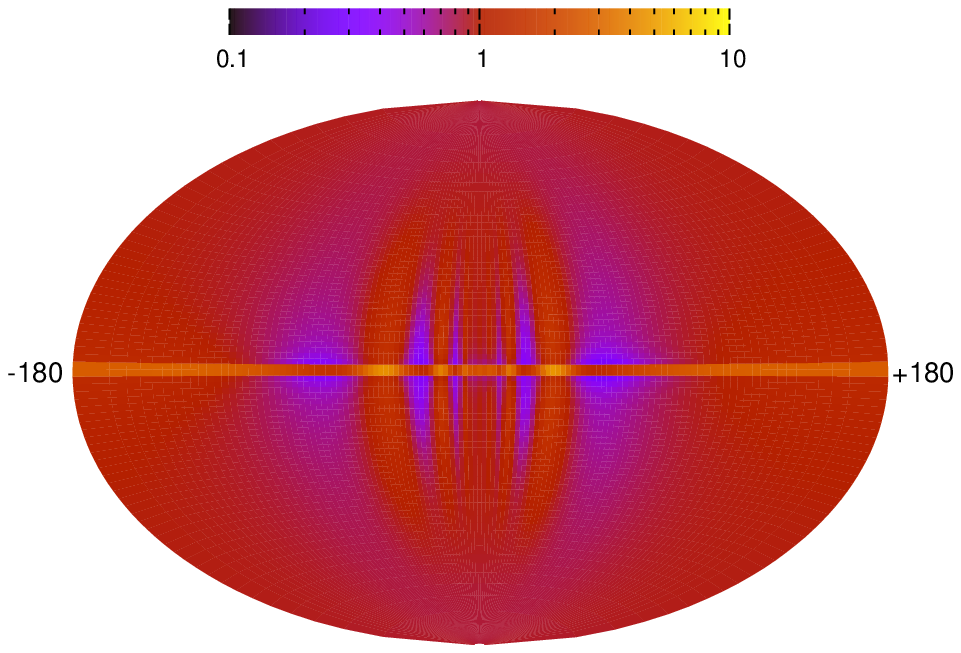,width=0.9\columnwidth}
\epsfig{file=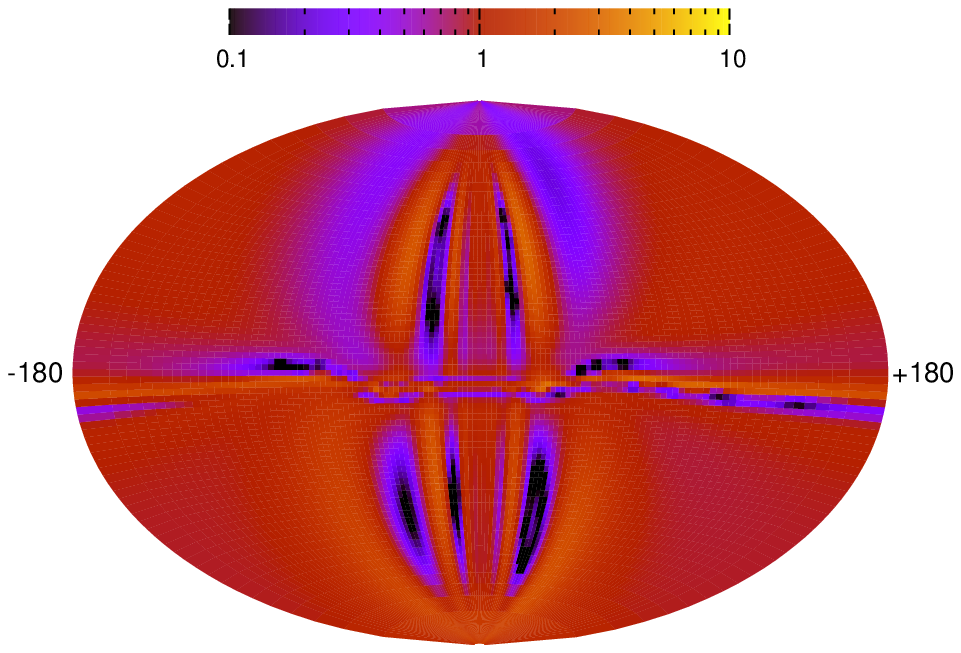,width=0.9\columnwidth}
\epsfig{file=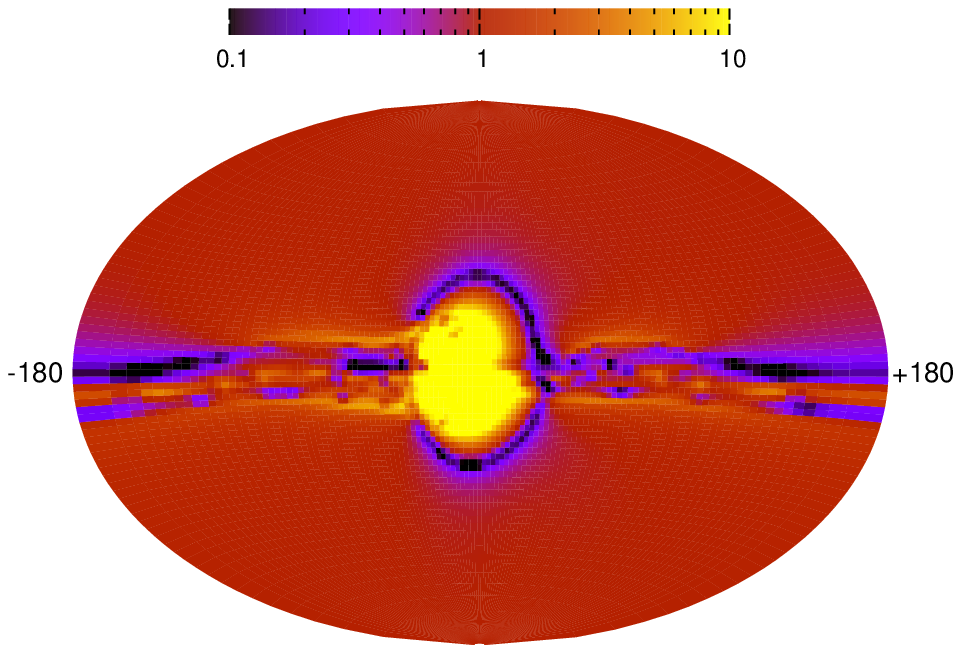,width=0.9\columnwidth} \caption{The
extragalactic ``exposure'' maps for the TT (top), HMR (middle), and
PS (bottom) model for a rigidity of 4$\times$ 10$^{19}$~V. All maps
use a Hammer-Aitoff projection of galactic coordinates.
\label{fig:lensing}}
\end{center}
\end{figure}
Apart for deflections, (de-)~focusing effects of the GMF effectively
modify the exposure to the extragalactic sky. This modification can
be calculated by back-tracking a large number of events and looking
at the obtained map of event numbers per solid angle outside the
Galaxy. For the purpose of illustration, we show in
Fig.~\ref{fig:lensing} some relative exposure maps obtained for
fixed rigidity for the three chosen GMF models. They were obtained
with the technique described in~\cite{Harari:1999it}, and
essentially represent the ratio
$\omega_B(l,b)=d\Omega_\infty(l',b')/d\Omega_\oplus(l,b)$, where
$d\Omega_\oplus$ is an infinitesimal small cone at Earth (around the
direction $l,b$) transported along the trajectory of a charged
particle to the border of the Galaxy $d\Omega_\infty$ (around the
new position $l'(l,b),b'(l,b)$). If $\omega_B(l,b)$ deviates
significantly from one, the corrected exposure has to be used in
(auto-)~correlation studies. Note how this effect is present in all
the models at least for cosmic rays observed at the Earth along the
Galactic plane.

A remark on the role of the turbulent GMF is in order. A comparison
of Eq.~(\ref{deltarms}) and Eq.~(\ref{regdeflEq}) shows that
deflections in the turbulent GMF are sub-leading. However, this does
not ensure automatically that magnetic lensing by the turbulent
fields is irrelevant at high energies, because lensing depends on
the gradient of the field and the critical energy for amplification
is proportional to $L_c^{-1/2}$. The detailed analysis
of~\cite{Harari:2002dy} showed indeed that small-scale turbulence
can produce relatively large magnification effects, and argued that
it may even be responsible for (some of) the multiplets seen by
AGASA above $4\times 10^{19}$~eV.
However, we expect that the random GMF introduces a small scale (and
strongly energy dependent) correction of the sky map on the top of
the magnification effects of the regular GMF. Then, we argue that
neglecting the turbulent GMF in our analysis is justified both for
intrinsic reasons and experimental limitations. First, possible
lensing effects by the random GMF are weakened by the presence of a
regular field component~\cite{Harari:2002dy}: Already without
regular field, the magnification peaks are quite narrow in energy
space, $\Delta E/E\sim 20\%$, and thus their width $\Delta E/E$ is
comparable to the energy resolution of CR experiments. The presence
of the regular GMF narrows these magnification peaks further by a
factor of several. Due to the limited energy resolutions of current
experiments, the turbulent lensing signature would be smeared out.
Second, the experimental angular resolution introduces an additional
averaging effect. Thus it is seems unlikely that lensing effects of
the turbulent GMF lead at present to distinctive effects taking into
account the current experimental limitations.
Finally, note that from a phenomenological point of view our
analysis is the same both if the multiplets are due to intrinsically
strong UHECRs sources or due to turbulent lensing.

\section{AGASA data sample}
\label{AGASAold}

In order to make the general considerations of the previous section
more concrete, we will discuss here some applications to the AGASA
data. The AGASA experiment has published the arrival directions of
their data until May 2000 with zenith angle $<45^\circ$ and energy
above $4\times 10^{19}$~eV~\cite{AGASA}. This data set consists of
$N=57$ CRs and contains a clustered component with four pairs and
one triplet within $2.5^\circ$ that has been
interpreted as first signature of point sources of UHECRs.
The reconstruction of the original CR arrival directions and the
estimate of their errors is obviously an important first step in the
identification of astrophysical CR sources.

\begin{figure}[!htb]
\begin{center}
\epsfig{file=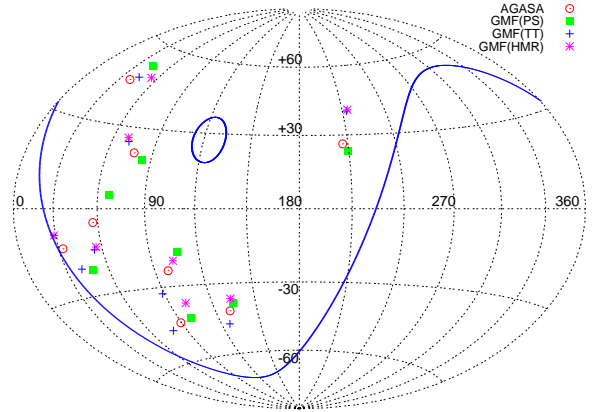,width=1.0\columnwidth} \caption{The
AGASA data set of CRs with estimated energy exceeding
  $10^{20}$~eV together with their reconstructed positions assuming
  Carbon primaries and the three GMF models discussed in the text.
  The line bounds the region visible to AGASA for zenith angles $\leq
  45^\circ$; note that for a better visibility, this map is centered
  around $l=180^\circ$, since AGASA is blind to the Galactic Center.
\label{directionsA}}
\end{center}
\end{figure}

In Fig.~\ref{directionsA}, we show the measured directions of all
CRs in the AGASA data with $E\geq 10^{20}$~eV together with their
reconstructed arrival direction at the border of the Galaxy for the
case of Carbon primaries. In the southern Galactic hemisphere, the
TT model often produces opposite deflections with respect to the HMR
and PS models, because of its antisymmetric field configuration. A
longitudinal shift is often appreciable in the PS model, as
consequence of the dipolar component we added. The larger deflection
found in the TT and HMR models for high latitudes is explained by
the stronger regular halo field these models use. The best chances
for source identification arise obviously by looking at directions
opposite to the GC. On the other hand, observations towards the GC
have a certain importance to use the UHECRs as diagnostic tool for
GMF, see e.g.~\cite{TTiy}.

Before turning to the statistical analysis, we briefly recall the
estimators we will use in the following. The autocorrelation
function $w_1$ is defined as
\be
 w_1 = \sum_{i=1}^{N_d}\sum_{j=1}^{i-1} \Theta(\ell-\ell_{ij}) \,,
\ee
where $\ell_{ij}$ is the angular distance between the two cosmic
rays $i$ and $j$, $\ell$ the chosen bin size, $\Theta$ the step
function, and $N_d$ is the number of CRs considered. Analogously,
one can define the correlation function $w_{\rm s}$ as
\be
 w_{\rm s} = \sum_{i=1}^{N_d}\sum_{a=1}^{N_{\rm s}} \Theta(\ell-\ell_{ia}) \,,
\ee
where $\ell_{ia}$ is the angular distance between the CR $i$ and the
candidate source $a$ and $N_{\rm s}$ is the number of source objects
considered.

\subsection{Autocorrelation analysis}
Let us  discuss now how the small-scale clustering observed by the
AGASA experiment is modified by the GMF. Note from
Fig.~\ref{fig:lensing} that even for protons the (de-)~magnification
of the exposure is already significant at energies $4\times
10^{19}$~eV. Neglecting the influence of the GMF (i.e., assuming
neutral primaries), one generates a large number of Monte Carlo sets
of CRs, each consisting of $N_d$ CRs distributed according to the
geometrical exposure $\omega_{\rm exp}$ of AGASA. The fraction of MC
sets that has a value of the first bin of the autocorrelation
function $w_1$ larger or equal to the observed one, $w_1^\ast$, is
called the chance probability $P$ of the signal. In the following,
we shall consider a bin-width of 2.5$^\circ$, suited for AGASA
angular resolution. For a nonzero GMF, one uses the back-tracking
method: the observed arrival directions on Earth are back-tracked
following a particle with the opposite charge to the boundary of the
GMF. Then the value $w$ of the autocorrelation function is
calculated. Since also the exposure is changed by the GMF, the CRs
of the Monte Carlo sets have to be generated now using as exposure
$\omega_{\rm tot}({\cal R},l,b) = \omega_{\rm exp}(l,b)
\omega_B({\cal R},l,b)$ (the energy spectrum is sampled according to
the spectrum observed by AGASA, ${\rm d}N/{\rm d}E\propto
E^{-2.7}$). This is automatically fulfilled if one back-tracks
uniformly distributed Monte Carlo sets in the same GMF\footnote{For
very large statistics, however, it is numerically more convenient to
explicitly calculate $\omega_B({\cal R},l,b)$ and generate the Monte
Carlo sets accordingly.}. The resulting chance probability is called
$P$ in Table~\ref{w1-p}. The error due to the statistical
fluctuation in the Monte Carlo sets is on the last digit reported.
For illustration, we show also the chance probability $P_0$
calculated using only the experimental exposure (or $\omega_B=1$)
that overestimates that clusters come indeed from the same source.

Correcting for the GMF reduces for all three GMF models the value of
$w_1$. 
While however for the TT and HMR models two doublets above
$5\times 10^{19}$~eV disappear, the PS model loses one low-energy
doublet. Thus, either some of the pairs are created by the focusing
effect of the GMF, or the GMF and especially its halo component is
not well enough reproduced by the models. In the latter case, ``true
pairs'' are destroyed by the the incorrect reconstruction of their
trajectories in the GMF. Reference~\cite{Troitsky:2005bc} discussed in
detail the effect of the GMF on the AGASA triplet and found that
current GMF models defocus it. If the clustering is
physical, this could be explained by a wrong modeling of the GMF in
that high-latitude region. Alternatively, our assumptions of
negligible deflections in
the extragalactic magnetic field and of protons as primaries could be
wrong. Note that the effect of the GMF induced exposure to the
extragalactic sky is not in general negligible. The fact that $P_0$
is only somewhat smaller than $P$ hints that only a small fraction
of clusters might be caused by magnetic lensing (in the regular
field). For the AGASA data set this is expected, because the field
of view of this experiment peaks away from the inner regions of the
Galactic plane, and the GC in particular.
Finally, we note that the energy threshold for which the chance of
clustering is minimal decreases for the TT model. This change is
however rather small and a larger data set is needed for any
definite conclusion.

In Table \ref{w1-Q} the same analysis is performed for the TT model
only, but assuming also $Z=+2,-1$. In no case is the improvement with 
respect to the $Z=0$ case appreciable.

\begin{table*}[!htb]
\begin{center}
\begin{tabular}{|c|c|cc|ccc|ccc|ccc|}
\hline
 \multicolumn{2}{|c|}{Model}& \multicolumn{2}{c|}{no GMF}&\multicolumn{3}{c|}{TT}&\multicolumn{3}{c|}{HMR}&\multicolumn{3}{c|}{PS}\\
\hline $E_{\min}/10^{19}$~eV & $N_d$ & $w_1$ & $P[\%]$ & $w_1$ & $P[\%]$ &  $P_0[\%]$  & $w_1$ & $P[\%]$ &  $P_0[\%]$  & $w_1$ & $P[\%]$ &  $P_0[\%]$\\
\hline
5.0 &   32 & 4  & 0.22  & 2 & 8.9 &  8.5  & 2  & 9.5   & 8.5 & 4  & 0.27   & 0.22  \\
4.5 &   43 & 6  & 0.05 & 4 & 1.5 &  1.4  & 4  & 1.8   & 1.4  & 6  & 0.10   & 0.05 \\
4.0 &   57 & 7  & 0.18  & 6 &0.8 & 0.7  & 5  & 3.3   & 2.4   & 6  & 1.09   & 0.72\\
\hline
\end{tabular}
\caption{ Number $N$ of CRs with energy $E\geq E_{\min}$ and zenith
angle $\theta\leq 45^\circ$; the values of the first bin of the
autocorrelation function $w_1$, and the chance probability
$P(w_1\geq w_1^\ast)$  from an isotropic test distribution are shown
for the two cases with ($P$) and without ($P_0$) correction of the
exposure, respectively. Proton primaries are assumed.
\label{w1-p}}
\end{center}
\end{table*}
\begin{table*}[!htb]
\begin{center}
\begin{tabular}{|c|c|cc|cc|cc|cc|}
\hline
 \multicolumn{2}{|c|}{Model}& \multicolumn{2}{c|}{$Z$=0}&\multicolumn{2}{c|}{$Z$=+1}&\multicolumn{2}{c|}{$Z$=+2}&\multicolumn{2}{c|}{$Z$=-1}\\
\hline $E_{\min}/10^{19}$~eV & $N_d$ & $w_1$ & $P[\%]$  & $w_1$ & $P[\%]$  & $w_1$ & $P[\%]$  & $w_1$ & $P[\%]$\\
\hline
5.0 &   32 & 4  & 0.22  & 2 & 8.9  & 1  & 41 & 4  & 0.27   \\
4.5 &   43 & 6  & 0.05 & 4 & 1.5   & 4  & 1.8 & 6  & 0.10    \\
4.0 &   57 & 7  & 0.18  & 6 &0.8  & 5  & 3.5   & 6  & 1.08   \\
\hline
\end{tabular}
\caption{As in Table \ref{w1-p}, but for different charges of the
primary (TT model). \label{w1-Q}}
\end{center}
\end{table*}
\subsection{Correlations with BL Lacs}\label{corrBLlac}

Tinyakov and Tkachev examined in~\cite{Tinyakov:2001ir} if
correlations of UHECRs arrival directions with BL Lacs improve after
correcting for deflections in the GMF (see also~\cite{otherBL}). The
significance of the correlation found is still debated, and we just
choose this example as an illustration how correlation of UHECR
arrival directions with sources can be used to test the GMF model
and the primary charge. We use from the BL Lac
catalogue~\cite{vcv11th} all confirmed BL Lacs with magnitude
smaller than 18 (187 objects in the whole sky).

\begin{table*}[!htb]
\begin{center}
\begin{tabular}{|c|c|cc|ccc|ccc|ccc|}
\hline
\multicolumn{2}{|c|}{Model}& \multicolumn{2}{c|}{no GMF}&\multicolumn{3}{c|}{TT}&\multicolumn{3}{c|}{HMR}&\multicolumn{3}{c|}{PS}\\
\hline $E_{\min}/10^{19}$~eV & $N_d$ & $w_{\rm bl}$ & $P[\%]$ & $w_{\rm bl}$ & $P[\%]$ &  $P_0[\%]$  & $w_{\rm bl}$ & $P[\%]$ &  $P_0[\%]$  & $w_{\rm bl}$ & $P[\%]$ &  $P_0[\%]$\\
\hline
5.0 &   32 & 8  & 8.8 &  8 & 9.7  &  8.8  &  4  & 65  & 64  & 7  & 17   & 17 \\
4.5 &   43 & 11 & 4.7 & 14 & 0.6  &  0.5  &  7  & 49  & 39  & 7  & 39   & 39 \\
4.0 &   57 & 13 & 6.6 & 20 & 0.05 &  0.05 & 11  & 20  & 18  & 7  & 67   & 67\\
\hline
\end{tabular}
\caption{As in Table \ref{w1-p}, but for the correlation function
and the BL Lac catalogue discussed in the text.
\label{tab:corrBL}}
\end{center}
\end{table*}

In Table~\ref{tab:corrBL}, we show the chance probability to observe
a stronger correlation taking into account the three different GMF
models and assuming proton primaries. An improvement of the
correlation signal is found only for the TT model, while for the two
other models the correlation becomes weaker. In
Table~\ref{tab:corrBLQ}, the same analysis is performed for $Z=+2$
and $-1$ using the TT model. An improvement with respect to the $Z=0$
case is only found for protons, i.e. for $Z=+1$.
Obviously, the two cases $Z=+2$ and -1, i.e. a CR flux consisting
only of Helium nuclei or anti-protons, is not expected
theoretically, and is used mainly to illustrate the potential of
this method to distinguish between different charges.
Although it is difficult to draw strong conclusions given present
uncertainties, this example clearly shows how UHECRs observations
might be used to constrain the GMF models and/or to determine the
charge of the CR primaries.

\begin{table*}[!htb]
\begin{center}
\begin{tabular}{|c|c|cc|cc|cc|cc|}
\hline
 \multicolumn{2}{|c|}{Model }& \multicolumn{2}{c|}{$Z$=0}&\multicolumn{2}{c|}{$Z$=+1}&\multicolumn{2}{c|}{$Z$=+2}&\multicolumn{2}{c|}{$Z$=-1}\\
\hline $E_{\min}/10^{19}$~eV & $N_d$ & $w_{\rm bl}$ & $P[\%]$  & $w_{\rm bl}$ & $P[\%]$  & $w_{\rm bl}$ & $P[\%]$
& $w_{\rm bl}$ & $P[\%]$ \\
\hline
5.0 &   32 & 8  & 8.8 &  8 & 9.7   & 7   & 20   & 6  & 27   \\
4.5 &   43 & 11 & 4.7 & 14 & 0.6   & 12  & 3.4  & 7  & 36    \\
4.0 &   57 & 13 & 6.6 & 20 & 0.05 & 13  & 9.9  & 8  & 50   \\
\hline
\end{tabular}
\caption{As in Table \ref{tab:corrBL}, but for different charges
of the primary (TT model).\label{tab:corrBLQ}}
\end{center}
\end{table*}

\section{Conclusion and Perspectives}\label{conc}

In this work we have discussed in detail the effect of the regular
component of the Galactic magnetic field on the propagation of
UHECR, using three GMF models discussed previously in the
literature. Both in small-scale clustering and correlation studies,
the GMF might be used as a sort of natural spectrograph for UHECR,
thus helping in identifying sources, restricting the GMF models as
well as the chemical composition of the primaries. Notice that the
latter point is an important prerequisite to use UHECR data to study
strong interaction at energy scales otherwise inaccessible to
laboratory experiments.

As a consequence of the existence of blind regions, we have argued
that in some GMF models the observed isotropy of cosmic ray flux
e.g. around a few 10$^{17}\,$eV might disfavor a transition to an
extragalactic flux of protons at too low energies, say below
10$^{17.5}\,$eV. Unfortunately the effect is model-dependent, and a
better characterization of the GMF is needed to draw more robust
conclusions.

At higher energies, the cosmic rays should enter the ballistic regime.
We have provided in tabular and pictorial form an estimate of
the typical deflection suffered by UHECRs in the small deflection
limit and of its variability from model to model.
These results imply that, if the angular resolution of current
experiments has to be fully exploited,
deflections in the
GMF cannot be neglected even for $E=10^{20}\,$eV protons,
especially for trajectories along the Galactic plane or crossing
the GC region. Since the magnitude as well as the direction of the
deflections are very model-dependent, it is difficult
to correct for deflections with the present knowledge about the GMF.

We have also emphasized that, to the purpose of statistical analyses
like auto-correlation/cross-correlation studies, the GMF can effectively
act in distorting the exposure of the experiment to the extragalactic sky,
and we provided some maps of this ``exposure-modification'' effect.
As a consequence, to estimate the chance probability
e.g. of small-scale clustering, one should take into account the
role of the GMF. We showed that this effect is already appreciable
in the data published by AGASA, although its field of view do not include the
central regions of the Galaxy.
Especially for experiments in the southern hemisphere like Auger,
one might wonder if excluding some part of the data from (auto-)
correlation studies might lead to more robust analyses, at least as
long as no reliable model for the GMF is established. We stress that
the required cuts would be quite drastic. For instance, fixing
$E_{\rm min}=4\times 10^{19}$ eV and considering only sky regions
where $|\omega_B-1|<0.2$ would exclude
\\{}\\
{\it TT:} $-5^\circ\,< b_G\,<\,5^\circ$ for all $l_G$ and
$-60^\circ\,<\, b_G<\,60^\circ$ for $-94^\circ<l_G<75^\circ$,\\
{\it PS:}  $-25^\circ< b_G<22^\circ$ for all $l_G$ and $-38^\circ<
b_G< 40^\circ$ for $-33^\circ<l_G<35^\circ$,\\
{\it HMR:} $-11^\circ
< b_G< 8^\circ$ for all $l_G$ and all $b_G$ for
$-90^\circ<l_G<90^\circ$.
\\{}\\
Note that in the HMR model one would cut more than half of the sky.
A more reasonable prescription is to evaluate the robustness of the
significance of autocorrelation and correlation studies by taking
into account several models of GMF and several primary charges, as
we have done in this paper for the public available AGASA catalogue.
Although the present statistics does not allow to draw strong
conclusions, we have not found any signal of improvement after the
correction for GMF. This could point to an insufficient knowledge of
the field or to a significantly heterogeneous chemical composition
of the primaries. Finally, the AGASA signal might only be a chance
fluctuation.

Independently on the outcome, performing statistical analyses taking
into account several models of GMF and several primary charges is an
useful exercise. In the most pessimistic case, it allows one to
quantify in an approximate way the contribution of the GMF to the
overall uncertainty in the chance probability of a candidate signal.
On the other hand, a strong improvement in the significance of a
statistical estimator might favor a certain GMF model and/or primary
charge assignment. For example, by repeating the study
of~\cite{Tinyakov:2001ir} we have found that the significant
correlation of BL Lacs with UHECRs is strongly dependent on the GMF
and primary adopted, and is present only in the TT model of the GMF
for $Z=1$. Although this evidence needs confirmation with a larger
data set of UHECRs, it may be the start of the era of UHECR
astronomy.

\section*{Acknowledgments}

We are grateful to Dmitry Semikoz for collaboration during the
initial phase of this work and for useful comments on the
manuscript. M.K.\ was partially supported by the Deutsche
Forschungsgemeinschaft within the Emmy-Noether program. PS acknowledges
the support by the Deut\-sche For\-schungs\-ge\-mein\-schaft under grant
SFB 375 and by the European Network of Theoretical Astroparticle Physics
ILIAS/N6 under contract number RII3-CT-2004-506222

\end{document}